\documentclass[12pt]{elsarticle}
\usepackage{epsfig}
\usepackage{amsthm}
\usepackage{amsmath}
\usepackage{amssymb}
\usepackage{amsfonts}
\usepackage{mathpazo}
\usepackage{times}
\usepackage{color}
\usepackage{hyperref}
\usepackage{tikz}
\usepackage[normalem]{ulem}
\usepackage[colorinlistoftodos,prependcaption,textsize=tiny]{todonotes}
\usetikzlibrary{decorations.pathmorphing,decorations.markings,automata,arrows,positioning,shapes,matrix}
\usetikzlibrary{arrows,chains,matrix,positioning,scopes,decorations,calc,shapes,decorations.pathreplacing}

\usetikzlibrary{arrows.meta,
	calc, chains,
	quotes,
	positioning,
	shapes.geometric}

\usepackage{refcount}
\usepackage{listings}

\newcounter{clm}

\newtheorem{theorem}{Theorem}
\newtheorem{proposition}[theorem]{Proposition}
\newtheorem{claim}[clm]{Claim}

\newtheorem{reptheorem}{Theorem}

\newdefinition{definition}[theorem]{Definition}
\newdefinition{conjecture}[theorem]{Conjecture}
\newdefinition{observation}[theorem]{Observation}
\newdefinition{example}[theorem]{Example}
\newdefinition{remark}[theorem]{Remark}
\newdefinition{corollary}[theorem]{Corollary}
\newdefinition{lemma}[theorem]{Lemma}

\DeclareMathOperator{\shift}{shift}

\makeatletter
\tikzset{reset preactions/.code={\def\tikz@preactions{}}}
\makeatother

\usepackage{amssymb}
\usepackage{algpseudocode,algorithm}
\usepackage{subfigure}

\newcommand{\val}{\mathit{val}}

\newcommand{\fneg}{\mathit{neg}}
\newcommand{\head}{\mathit{head}}

\usepackage{algpseudocode,algorithm,algorithmicx}

\algdef{SE}[DOWHILE]{Do}{doWhile}{\algorithmicdo}[1]{\algorithmicwhile\textsl{}\ #1}%

\renewcommand{\leq}{\leqslant}
\renewcommand{\geq}{\geqslant}

\begin {document}

\begin{frontmatter}
\title{De Bruijn Sequences: From Games to Shift-Rules to a Proof of the  Fredricksen-Kessler-Maiorana Theorem}


	\author{Gal Amram}
	\ead{galamra@cs.bgu.ac.il}

	\author{Amir Rubin}
	\ead{amirrub@cs.bgu.ac.il}

	\author{Yotam Svoray}
	\ead{ysavorai@post.bgu.ac.il}

	\author{Gera Weiss}
	\ead{geraw@cs.bgu.ac.il}

	\address{Department of Computer Science, Ben-Gurion University of The Negev}

	\begin{abstract}
    We present a combinatorial game and propose efficiently computable optimal strategies. We then show how these strategies can be translated to efficiently computable shift-rules for the well known prefer-max and prefer-min De Bruijn sequences, in both forward and backward directions. Using these shift-rules, we provide a new proof of the well known theorem by Fredricksen, Kessler, and Maiorana on De Bruijn sequences and Lyndon words.
	\end{abstract}
\end{frontmatter}

\section{Introduction}
A De Bruijn sequence of order $n$ on the alphabet $[k]=\{0,\dots,k-1\}$ is a cyclic sequence of length $k^n$ such that every possible word of length $n$ over this alphabet appears exactly once as a subword~\cite{bruijn1946combinatorial}. In this work we focus on two of the most famous De Bruijn sequences called the prefer-max and the prefer-min sequences~\cite{martin1934problem, fredricksen1977lexicographic} obtained by starting with $0^n$, for the prefer-max, or $(k-1)^n$, for the prefer-min, and adding to each prefix the maximal/minimal value in $[k]$ such that the suffix of length $n$ does not appear as a subword of the prefix. 

A shift-rule  of a De Bruijn sequence is a mapping $\shift \colon [k]^n \to [k]^n$ 
such that, for each word $w = \sigma_1 \dots \sigma_n$, $\shift(w)$ is the word $\sigma_2 \dots \sigma_n \tau$  where $\tau$ is the symbol that follows $w$ in the sequence (for the last word $\tau$ is the  first symbol in the sequence). Amram et al.~\cite{amram2017efficient, DBLP:journals/dm/AmramR20} proposed an efficiently computable\footnote{In this paper the term ``efficiently computable" always means $O(n)$ time and space} shift-rule for the sequence defined by the concatenation of an ordered list of words called Lyndon words. 
They then conclude that their shift-rule generates the prefer-min sequence, using the well known FKM theorem by Fredricksen, Kessler, and Miorana~\cite{fredricksen1972generation, fredricksen1977lexicographic} which says that this concatenation is in fact the prefer-min sequence.

When $k=2$, the prefer-max sequence is called prefer-one, for obvious reasons. Weiss~\cite{weiss2007combinatorial} proposed a combinatorial game such that a play of two optimal players yields the prefer-one sequence. He also developed efficiently computable optimal strategies for both players in this game and used these to propose an efficiently computable shift rule for the prefer-one sequence.

The first result in this paper is a generalization of Weiss's result to larger alphabets. Specifically, we present a two-player combinatorial game over arbitrary alphabet ($k \geq 2$) such that if both players play optimally the play of the game gives the prefer-max sequence. Independently, DiMuro~\cite{dimuro2018warden} published a text that describes the same game in a slightly different way, named ``The Warden Game''. Note that DiMuro did not present an efficiently computable strategies for the game, thus, his work does not directly give an efficient algorithm for generating the sequence.

Our second result is efficiently computable optimal strategies for both players in the new game. These strategies are a generalization of the strategies proposed by Weiss.

The third result in this paper is the development of efficiently computable optimal strategies which can be used to construct an efficiently computable shift-rules for the prefer-max and prefer-min sequences in both forward and backward directions. Note that reversing the direction is not an issue in the binary case, since one can simply try both options, but it is an issue in the general case since trying all options adds a factor of $k$ to the complexity. 

Finally, we show that our shift-rule is equivalent to the one presented by Amram et al.~\cite{amram2017efficient} and use this fact to prove the FKM theorem. This is straight-forward since our shift-rule generates the prefer-min sequence directly while Amram's shift-rule generates the sequence produced by the concatenation of the Lyndon words.

An implementation of this work can be found online.\footnote{\url{https://github.com/amirubin87/De-Bruijn-Sequences}}

\section{Preliminaries}


The directed De Bruijn graph of order $n$ over the alphabet $[k]$ is the graph whose vertices are the words of length $n$ over the alphabet $[k]$ (i.e. the set $[k]^n$) and whose edges are such that each vertex $v=\tau x$ is connected with directed edges to all vertices in $\{ x \sigma \colon \sigma \in [k] \}$. 

There is a one-to-one correspondence between De Bruijn sequences and Hamiltonian cycles in the De Bruijn graph of the same order and alphabet, described in~\cite{bruijn1946combinatorial} and it is as follows:
\begin{enumerate}
	\item If, for each $i$, $w_i=x_i \sigma_i$, and  $(w_1, w_2, \dots)$ is an Hamiltonian cycle then $(\sigma_1, \sigma_2,\dots)$ is a De Bruijn sequence.

	\item A Hamiltonian cycle can be constructed from a De Bruijn sequence $(\sigma_1,\dots,\sigma_{k^n})$ 
	by visiting the vertex ${\sigma_1 \cdots \sigma_n}$, then ${\sigma_2 \cdots \sigma_{n+1}}$ and so on, until we return to where we started.
\end{enumerate}

In this paper we focus on a specific Hamiltonian cycle in the De Bruijn graph called the prefer-max cycle (and the corresponding prefer-max De Bruijn sequence) defined as follows.

\begin{definition} \label{def:pref-max}
	The $(n,k)$-prefer-max cycle, $(w_i)_{i=0}^{k^n-1}$, is defined by $w_0=0^{n-1}(k-1)$ and if $w_i=\tau x$ then $w_{i+1}= x\sigma$ where $\sigma$ is the maximal letter such that $x \sigma \notin \{w_0,\dots,w_{i}\}$. We denote $w_i \prec w_j$ if $i<j$ in this sequence.
	
\end{definition}

We also consider the $(n,k)$-prefer-min cycle defined in a similar way, by starting with $w_0=(k-1)^{n-1}0$ and using the minimal $\tau$ instead of the maximal.

Martin~\cite{Mar34} proved that the cycle given in Definition~\ref{def:pref-max} is Hamiltonian, i.e., that for $k^n-1$ steps there is always a $\sigma$ such that $w \sigma \notin \{w_0,\dots,w_{i}\}$. This means that $w_{i+1}$ is well defined. Definition~\ref{def:pref-max} above is demonstrated in Example~\ref{example:pref-max}.

\begin{example}\label{example:pref-max}
	For example, if we set $n=3$ and $k=3$, we have:
	$002 \to 022 \to 222 \to 221 \to 212 \to 122 \to 220 \to 202 \to 021 \to 211 \to 112 \to 121 \to 210 \to 102 \to 020    \to 201 \to 012 \to 120 \to 200 \to 001 \to 011 \to 111 \to 110 \to 101 \to 010 \to 100 \to 000$.
\end{example}



\section{A useful property of the prefer max sequence}

Before diving to the specific contribution of this paper, we identify a useful property of the prefer-max cycle, as follows.

\begin{lemma}\label{lemma:xy}
	For any words $x,y$ such that $|x|+|y|=n-1$,
    let $(\sigma_i)_{i=0}^{k-1}$ be such that  $(x \sigma_i y)_{i=0}^{k-1}$ is the subsequence of the  $(n,k)$-prefer-max cycle 
    consisting of all the words that begin with $x$ and end with $y$.
   	Then, there exist some $d \in [k]$ such that the sequence is sorted with the exception of $\sigma_d=0$: $$(\sigma_i)_{i=0}^{k-1} =(k-1,\dots, k-d, 0, k-d-1, \dots, 1).$$ 

\end{lemma}
\begin{proof}
    
	By induction on the length of $y$. If $y$ is the empty word, the statement is true (with $d=k-1$) from the definition of the prefer-max cycle (Definition~\ref{def:pref-max}). For the induction step, assume that for some $t<n-1$ the statement is true for all $y$ of length $t$. 
    We need to show that $x\sigma_2y\tau \prec x\sigma_1y\tau $ for any symbols $\tau$ , $0<\sigma_1 <\sigma_2$ and any word $x$ of length $n-t-2$.  Let $v_1=x\sigma_1y\tau$, and $v_2=x\sigma_2y\tau$.
    As $\sigma_1 \neq 0$, for each $\psi\in[k]$, $x\sigma_1y\psi$ is not the first element in the cycle, thus, it has a predecessor.
    By the definition of the prefer-max cycle  $x \sigma_1 y (k-1) \prec x \sigma_1 y (k-2)\prec \cdots \prec x \sigma_1 y 0$. Because the
    predecessor of each of these vertices is in  $\{\varphi x \sigma_1 y \colon  \varphi \in [k]\}$ we have that $k-\tau$ vertices in this set precede $v_1$.
    By the induction hypothesis we get that $\varphi x \sigma_2 y \prec \varphi x \sigma_1 y$, for any $\varphi$, and therefore at least  $k-\tau$ vertices in $\{\varphi x \sigma_2 y \colon  \varphi \in [k]\}$  precede $v_1$. 
    The follower of each of these vertices is in  $\{x \sigma_2 y (k-1),x \sigma_2 y (k-2), \dots,x \sigma_2 y 0\}$ whose members, by definition, appear in decreasing lexicographical order in the prefer-max cycle. Therefore, at least $k-\tau$ vertices in $\{x \sigma_2 y (k-1),x \sigma_2 y (k-2), \dots,x \sigma_2 y 0\}$  precede $x \sigma_1 y \tau$.
	From this we get that $x \sigma_2 y \tau$ must be before $x \sigma_1 y \tau$.              
\end{proof}

\begin{example}\label{example:prop}
	Consider the sequence from Example~\ref{example:pref-max}, and choose $x=2$ and $y=0$. For these values, we get the subsequence: $(002, 022, 012)$ that gives us $(0,2,1)$ which is sorted if we remove 0 from it. 
\end{example}

\begin{observation}\label{obs:s1_s2}
For any $x \in [k]^{n-1}$, let $S_1=(x(k-1),\dots,x\sigma,\dots,x0)$ be the subsequence of $\{w_i\}_{i=0}^{k^n-1}$ of the words that start with $x$. By Lemma~\ref{lemma:xy}, there is a  $d\in[k]$ such that $S_2=( (k-1)x, \dots (k-d) x,0x, (k-d-1)x,\dots,1x )$  is a subsequence of $\{w_i\}_{i=0}^{k^n-1}$. Hence, the $i$th element of $S_1$ is preceded by the $i$th element of $S_2$, and every word $w_i = x\sigma, i \neq 0$, is preceded by $w_{i-1} \in \{\sigma x, 0 x, (\sigma+1) x\}$.
\end{observation}

Based on this observation, the game that we will present in the next section will focus on following the sequence backwards, specifically, on choosing which of the three possible predecessors of a state is chosen.  

\if{0}
        	             $? x \sigma_1 y$, $x\sigma_1 y (k-1)$,..., $? x \sigma_1 y$,$x\sigma_1 y(k-2)$,...,  $? x \sigma_1 y, x\sigma_1 y(k-3)$, ...,   $? x \sigma_1 y$,$x \sigma_1 y \tau$ 
                        /                                            /                                       /                                    /
           $? x \sigma_2 y$, $x\sigma_2 y(k-1)$,..., $? x \sigma_2 y$,$x\sigma_2 y(k-2)$,...,  $? x \sigma_2 y, x\sigma_2 y(k-3)$, ...,   $? x \sigma_2 y$,$x \sigma_2 y \tau$           
\fi           

\section{A combinatorial game for the prefer-max sequence }

The main object that we will analyse in this paper is the following combinatorial game, played between two players, Alice and Bob. A \emph{state} of the game is a word $s\in[k]^n$, and the \emph{initial-state} is $0^n$. In each game round, Bob plays first. If the state is $s=w\sigma$, Bob can either set the next state to be $(\sigma+1)w$, or pass control to Alice. Note that if $\sigma=k-1$  Bob cannot increase $\sigma$ and thus he must pass control to Alice. In case that Bob passes control, Alice gets to choose the next state. She, then, has two options: she can choose the successor state to be either $\sigma w$, or $0w$. 

Bob's goal is to reach an already-seen state $s\neq 0^n$. Alice's goal is to reach $0^n$ quickly. A play ends in a tie if $0^n$ is reached only after traversing all $k^n$-possible states. The next definition formalizes these requirements, and define the notions of strategies, state-progressions and plays.

\begin{definition}\label{def:shift-game}
The $(n,k)$-shift-game is a two-player combinatorial game defined as follows:
\begin{itemize}
 
    \item  \emph{Strategies} for the players Alice and Bob, respectively, are functions $A,B \colon [k]^n \to \{0,1\}$  such that $B(x(k-1))=A(x0)=0$ for all $x \in [k]^{n-1}$.\footnote{The numbers $1$ and $0$ that $A$ and $B$ assign to  states represent active and passive actions, respectively. Bob's active action is to increase the last symbol by $1$, and his passive action is to pass the turn to Alice. Alice's active action is to write $0$, wheres her passive action is to copy the last symbol.} 

    \item  A \emph{state progression} of the game with the strategies $A$ and $B$ is a (finite or infinite) sequence $s_0, s_1, \dots$ for some $s_0 \in [k]^n$ and if $s_t=x\sigma$ for $x\in[k]^{n-1}$ and $\sigma \in [k]$,  then $s_t$ is last in the sequence or:
    	$$s_{t+1}=\begin{cases}
    			(\sigma+1)x &
    			\text{if $B(s_t)=1$},                \\
    			0x          &
    			\text{if $B(s_t)=0$ and $A(s_t)=1$}, \\
    			\sigma x    &
    			\text{otherwise;}
    		\end{cases}$$

    \item A (complete) play of the $(n,k)$-shift-game is a state progression starting with $s_0=0^n$ and ending with $s_m \in \{s_0,\dots,s_{m-1}\}$, such that no prefix of it is a play.
    \item  Alice wins a play $s_0,\dots,s_m$ if $m< k^n$ and $s_m = 0^n$.
    \item  Bob wins a play  $s_0,\dots,s_m$ if $s_m \neq 0^n$.
    \item  A play  $s_0,\dots,s_m$ is a tie if  $m=k^n$ and $s_m=0^n$ .
\end{itemize}
\end{definition}

Note that the condition that $A(x0) =0$ in the first bullet is, in some sense, vacuous since $A(x0)=1$ will produce the same output. We added this requirement to assure uniqueness of strategies. Later, it will be convenient that the choice between active and passive actions is always meaningful.

The fact that not all functions are strategies is  not a problem because we can construct new strategies from existing ones using the following fact:
\begin{observation}\label{obs:strategyC}
If $S$ is a strategy for Alice or for Bob and $S'\colon [k]^n \to \{0,1\}$ is a function such that $S'(w) \leq S(w)$ for all $w$, then $S'$ is also a strategy for the same player.
\end{observation}

This game is a generalization of a game defined and analyzed in~\cite{weiss2007combinatorial}. There, the game was only defined for a binary alphabet ($k=2$) and was used to produce an efficient algorithm to construct the prefer-prefer-one sequence. Here, we show how the definition can be extended to a larger alphabet and how this can be used to produce an efficient algorithm for constructing any prefer-max sequence, for any $k \geq 2$. 

The game that we are proposing was independently proposed by DiMuro in~\cite{dimuro2018warden}. Compared to Dimuro's paper, the contribution of this paper is an efficiently computable winning strategy for each of the players and a method for using these strategies for efficiently computing a shift rule for De Bruijn sequences.

\section{Non-losing strategies for both players}\label{sec:game}

Next, we turn to establishing the connection between the prefer-max cycle and the shift-game. We will first define a pair of  strategies $A^*$ and $B^*$ such that if both Alice and Bob use the respective strategies, the play of the game follows the prefer-max cycle in reversed order. Then, we will show that $A^*$ and $B^*$ are the unique non-losing strategies for Alice and for Bob respectively. This will give us that the efficient implementations of non-losing strategies for both of the players, that we will provide in the following sections, can serve as an efficient shift-rule for the cycle. 

The strategies $A^*$ and $B^*$ use the prefer-max cycle as an internal `oracle', as specified in the following definition:

\begin{definition}\label{def:star}
	Considering $(w_i)_{i=0}^{k^n-1}$ from Definition~\ref{def:pref-max}, let $A^{*},B^{*} \colon [k]^n \to \{0,1\}$ be the strategies for Alice and Bob, respectively, defined by
	$$
		B^{*}(w_i) =
		\begin{cases}
			1 & \text{if $w_i=x\sigma \wedge w_{i-1}=(\sigma+1) x$}, \\
			0 & \text{otherwise;}
		\end{cases}
	$$
	
	and
	$$
		A^{*}(w_i) =
		\begin{cases}
			1 & \text{if $w_i=x\sigma \wedge \sigma \neq 0 \wedge  (w_{i-1}=0x \vee B^{*}(w_i) = 1)$}, \\
			0 & \text{otherwise}.
		\end{cases}
	$$

\end{definition}

From Definition~\ref{def:pref-max}, we can see that $A^*$ and $B^*$ are strategies: (1) By definition, $A^*(x0)=0$ for all $x \in [k]^{n-1}$;  (2) If $w_i=x(k-1)$ then $w_{i+1}$ cannot be $kx$ because this word is not in $[k]^n$, thus $B^*(w_i)$ must be zero. 

The essence of both strategies is to follow the prefer-max cycle. $B^*$ maps an element $x\sigma$ on the cycle to 1 if its predecessor is $(\sigma+1) x$. This means that Bob will force the next state to be $(\sigma+1)x$ in these cases and, by the rules of the game Alice will not be able to object.  In all other cases, $A^*$ maps $x\sigma$ to 1 when its predecessor is $0x$. If $\sigma$ is zero, there is no difference to the state of the game if Alice chooses to play or not, and therefore the condition $\sigma\neq 0$  in the definition of $A^*$ serves the technical role of forcing $A^*$ to be a strategy. Additionally, the condition that $A^*(w_i)=1$ when $B^*(w_i)=1$ was added to make it a winning strategy against any strategy that Bob may choose, as we will show below.

\begin{example}
	For example, for $n=3$ and $k=2$, if both players play according to the strategies above, then the resulting play is $000 \to 100 \to 010 \to 101 \to 110 \to 111 \to 011 \to 001 \to 000$ which yields a tie.
\end{example}

We can see that in the example above, the play is exactly the prefer-max cycle in reverse. 
In general, it is clear from Definition~\ref{def:star} that this is true for all $n$ and $k$, as stated in the next observation:

\begin{observation}\label{observation:SisW}
	Let $(s_t)_{t=0}^{m}$ be the play of the $(n,k)$-shift-game when Alice uses the $A^{*}$ strategy and Bob uses the $B^{*}$ strategy. Then,
	$(s_t)_{t=0}^{m-1} = (w_{k^n-t})_{t=1}^{k^n}$ where $(w_i)_{i=0}^{k^n-1}$ is the prefer-max cycle given in Definition~\ref{def:pref-max}. 
\end{observation}

The next proposition shows that the computation of $B^*$ can be reduced to a computation of $A^*$ over a slightly alternated input. We will use this fact to focus only on Alice's strategy, i.e., we'll develop an efficient algorithm to compute $A^*$ and, by the reduction specified in the following proposition, get the same for $B^*$.

\begin{proposition}\label{proposition:pmaxeq}
	$B^{*}(x\sigma)=A^{*}(x(\sigma+1))$ for every $x\in [k]^{n-1}$ and $\sigma<k-1$.
\end{proposition}
\begin{proof}
	Let $d \colon [k]^{n-1} \to [k]$ be defined by  $d(x)=|\{\tau \colon \tau x \prec 0x \}|$.
	From Lemma~\ref{lemma:xy} we get that for every $x \in [k]^{n-1}$, and $w_i=x \sigma$:
	\begin{itemize}
	    \item if $\sigma<k-d(x)-1$ then $w_{i-1}=(\sigma+1)x$;
	    \item if $\sigma=k-d(x)-1$ then $w_{i-1}=0x$;
	    \item if $\sigma>k-d(x)-1$ then $w_{i-1}=\sigma x$.
	\end{itemize} 
	Therefore, by the definition of $B^*$, $(B^{*}(x\sigma)=1) \Leftrightarrow  (\sigma < k-d(x)-1)$, and, by the definition of $A^*$, for every $\sigma>0$,  $(A^{*}(x\sigma)=1) \Leftrightarrow  (\sigma \leq k-d(x)-1)$. This means that $B^{*}(x\sigma)=1$ if and only if $A^{*}(x(\sigma+1))=1$.
\end{proof}

From Observation~\ref{observation:SisW} we get that if Alice plays according to $A^*$ and Bob plays according to $B^*$, the game ends with a tie.
Our next goal, in propositions \ref{proposition:Astar} and \ref{proposition:AUnique}, is to  show that $A^*$ is the only strategy that wins against any other strategy. Towards this goal, in Lemma~\ref{lemma:followers}, we first 
analyze the options for Bob and Alice in a given state $w_i=x\sigma$. Namely, we consider the state $w_{i-1}$ , that by Observation~\ref{observation:SisW} is the next state in the play of the game when $A^*$ and $B^*$ are used, and compare its position in the sequence relative to the other two options in $\{\sigma x, 0x, (\sigma+1)x\}$.

\begin{lemma} \label{lemma:followers}
Let $(w_i)_{i=0}^{k^n-1}$ be the $(n,k)$-prefer-max cycle. If $i>0$, $w_{i}=x\sigma$, $x\in [k]^{n-1}$, and $\sigma \in [k]$ then: 

\begin{enumerate}
\item $w_{i-1} \in \{\sigma x, 0 x , (\sigma+1) x \}$;

\item if $w_{i-1} = (\sigma+1) x $ then 
    $0 x \prec (\sigma+1) x $. If, in addition, $\sigma \neq 0$ then  $(\sigma+1) x \prec \sigma x$;

\item if $w_{i-1} = \tau x$ for $\tau \in \{0,\sigma\}$, then:
    \begin{itemize}
        \item if $\sigma \neq 0$, then $\tau x \prec \hat{\tau} x$ where $\hat{\tau} = \begin{cases} \sigma &\mbox{if } \tau=0;\\
    0 & \mbox{if } \tau=\sigma\end{cases}$
        
     
        \item if $\sigma < k-1$, then $ (\sigma+1) x \prec \tau x$ .

    \end{itemize} 
\end{enumerate}
\end{lemma}

\begin{proof}  
Item 1 is Observation~\ref{obs:s1_s2}.
To prove 2 and 3, we consider the two subsequences of the prefer-max cycle presented in Observation~\ref{obs:s1_s2}: $S_1=(x(k-1)),\dots,x\sigma,\dots,x0)$ and $S_2=( (k-1)x, \dots (k-d) x,0x, (k-d-1)x,\dots,1x )$. The $i$th element of $S_1$ is preceded by the $i$th element of $S_2$. Therefore, (1) if $\sigma >k-d-1$, $x\sigma$ is preceded by $\sigma x$; (2) if $\sigma=k-d-1$, $x\sigma$ is preceded by $0x$; (3) and otherwise if $\sigma<k-d-1$, then $x\sigma$ is preceded by $(\sigma+1)x$. 
The first inequality in 2 follows from cases (2) and (3) and the second inequality is a direct result of Lemma~\ref{lemma:xy}.

For the first bullet in 3, assume $\sigma \neq 0$. If $\tau=\sigma$, then $\sigma>k-d-1$ thus $\tau x=\sigma x \prec 0x = \hat{\tau}x$.
Similarly, if $\tau=0 $, then, as mentioned in case (2) above, $\sigma=k-d-1$. Following subsequence $S_2$ we have that $\tau x= 0x \prec (k-d-1)x =\sigma x=\hat{\tau}x$.

Lastly, to prove the second bullet in 3, assume $w_{i-1} \neq (\sigma+1) x$ and $\sigma < k-1$. Therefore, either (1) or (2) holds and thus $\sigma\geq k-d-1$. 
If $\sigma>k-d-1$ then by (1) $\tau=\sigma$ and $(\sigma+1)x\prec\sigma x$. Otherwise, if $\sigma=k-d-1$, then by (2) we have that $\tau=0$. By the definition of $d$, $(\sigma+1) x=(k-d)x\prec 0x$.
\end{proof}

The next proposition shows that we achieved the purpose of the game: the strategies that generate the prefer-max cycle as a play, $A^*$ and $B^*$, are optimal strategies for both players:

\begin{proposition}\label{proposition:Astar}
	If Alice applies the strategy $A^{*}$ she wins against any strategy that Bob may apply which is not $B^*$ and gets a tie against $B^*$.
\end{proposition}

\begin{proof}

First, from Observation~\ref{observation:SisW} we know that if the players play by the strategies $A^*$ and $B^*$ respectively we get a tie.  

Second, let $B\neq B^*$ be a strategy played by Bob and $i$  be such that $B(w_i)\neq B^*(w_i)$ and $w_i=x\sigma$: Let $w_t$ be the state that follows $w_i$ in the game played using $A^*$ and $B$.
We show, using the case analysis below, that $t < i-1$:

	\begin{itemize}
	    \item When $B(w_i)=0$, $B^*(w_i)=1$ and $A^*(w_i)=1$.
	         By the definition of $B^*$ we have that $w_{i-1}=(\sigma +1)x$.
	         By the definition of the game we have that $w_{t}=0x$. 
	         By Lemma~\ref{lemma:followers} (item 2), we have that $w_t=0x \prec (\sigma +1)x=w_{i-1}$ which gives us that $t < i-1$.
	         
	    \item When $B(w_i)=0$, $B^*(w_i)=1$ and $A^*(w_i)=0$.
	        By the definition of $B^*$ we have that $w_{i-1}=(\sigma +1)x$.
	        By the definition of $A^*$ we have that $\sigma = 0$, and so, by the definition of the game  $w_{t}=0x$.  
	         Again, by Lemma~\ref{lemma:followers} (item 2), we have that $w_t=0x \prec (\sigma +1)x=w_{i-1}$ which gives us that $t < i-1$.
	    \item When $B(w_i)=1$, $B^*(w_i)=0$ and $A^*(w_i)=1$.
	         By the definition of $A^*$ and $B^*$ we have that $w_{i-1}=0x$.
	         As $B$ is a strategy, $\sigma < k-1$.
	         By the definition of the game $w_t=(\sigma+1)x$.
	         By Lemma~\ref{lemma:followers} (item 3, second bullet, $\tau=0$), we have that $w_t = (\sigma+1)x \prec 0x = w_{i-1}$ which gives us that $t < i-1$.
	    
	    \item When $B(w_i)=1$, $B^*(w_i)=0$ and $A^*(w_i)=0$.
	         By the definition of $A^*$ and $B^*$ we have that $w_{i-1}=\sigma x$.
	         As $B$ is a strategy, $\sigma < k-1$.
	         By the definition of the game $w_t=(\sigma+1)x$.
	         By Lemma~\ref{lemma:followers} (item 3, second bullet, $\tau=\sigma$ ), we have that $w_t = (\sigma+1)x \prec \sigma x = w_{i-1}$ which gives us that $t < i-1$.         
	\end{itemize}

Together with the fact that $t=i-1$ if $B(w_i)=B^*(w_i)$, we get that $A^*$ and $B$ produce a play that is a strict subsequnce of the prefer-max cycle. In particular, this play ends in the state $s=0^n$ and it length is shorter than $k^n$ states, i.e., Alice wins.
\end{proof}

A direct result of the previous proposition is that $B^*$ is the unique non-losing strategy for Bob. We next show the same for $A^*$:

\begin{proposition}\label{proposition:AUnique}
	$A^*$ is the only non-losing strategy for Alice. 
\end{proposition}
\begin{proof}
We will show that for any strategy $A \neq A^*$ there is some strategy $B$ such that $B$ wins against $A$.


Let $B$ be defined by 
	$$
		B(w_i) =
		\begin{cases}
			B^*(w_i) & \text{if $A(w_i)=A^*(w_i)$}, \\
			0 & \text{otherwise.}
		\end{cases}
	$$
	
	Because  $B(w_i) \leq B^*(w_i)$, for any $w_i$, it is a strategy for Bob by Observation~\ref{obs:strategyC}.
	Let $i$ be the maximal integer such that $A(w_i)\neq A^*(w_i)$. By Observation~\ref{observation:SisW}, $A$ and $B$ produce a play with the prefix $w_{k^n-1},\dots,w_i$. We focus on the step following this prefix and argue that the next state in that play is some $w_t$ where $t \geq i$, in which case Bob wins.
    
    Write $w_i=x\sigma$ and note that, since $A(x\sigma)\neq A^*(x\sigma)$ we have that $\sigma\neq 0$. Write $w_{i-1}=\tau x$. We distinguish between two cases. 

\begin{enumerate}
\item  If $\tau\in\{0,\sigma\}$ then $w_t=\hat{\tau}x$ where $\hat{\tau} = |\tau-\sigma|$ and, by Lemma~\ref{lemma:followers}, item 3, first bullet, $w_{i-1} = \tau x \prec \hat{\tau}x = w_t$.

\item $\tau=\sigma+1$. 
By definition $A^*(w_i)=1$ and thus $A(w_i)=0$. Hence, the next state is $w_t = \sigma x$, and Lemma~\ref{lemma:followers} , item 2 gives us that $w_{i-1} = (\sigma+1)x \prec  \sigma x = w_t$.
\end{enumerate}

Now, to complete the proof, we show that $w^t\neq 0^n$. $w^t=0^n$ can only occur if $w_i=0^n\sigma$. First, consider the case in which $\sigma=k-1$. In this case, $w_{i-1}=0^n$ and thus $A^*((0^{n-1}(k-1))=1$. Hence, $A(0^{n-1}(k-1))=0$ and thus $w_t=(k-1)0^{n-1}\neq 0^n$. 

We turn to deal with the general case for $w_i=0^{n-1}\sigma$, where $\sigma\neq k-1$ (and also recall that $\sigma\neq 0$). Assume towards contradiction that $w_t = 0^n$. 

Since $0^n$ comes last in the prefer max cycle, by Lemma~\ref{lemma:xy}, the following two are subsequences of the prefer-max cycle:
\begin{itemize}
    \item $(0^{n-1}(k-1),\cdots,0^{n-1}1,0^{n-1}0)$ 
    \item $((k-1)0^{n-1},\cdots, 10^{n-1},00^{n-1})$
\end{itemize}
Because the element $0^{n-1}(k-1)$ is the first in the sequence it has no predecessor. The predecessor of any other element in the first subsequence, $w = 0^{n-1}\sigma$ is $(\sigma+1)0^{n-1}$, a member of the second subsequence. Specifically, this is true for $w_i$, and so $A^*(w_i)=1$ by the definition of $A^*$. Thus $A(w_i)=0$ which contradicts the fact that $w_t = 0^n$.

We showed that state $w_t \in \{w_{k^n-1},\dots,w_i\}$. Consequently, strategies $A$ and $B$ produce the play: $(w_{k^n-1},\dots, w_i, w_t)$ , $w_t \neq 0^n$ , in which Bob wins.
\end{proof}

In the next section we will develop an efficiently computable rule for a non-loosing strategy,  $A^\dagger$, for Alice. Using the above uniqueness property of $A^*$, we will conclude that $A^\dagger=A^*$, i.e., that we can compute $A^*$ efficiently.

\section{The $A^\dagger$ strategy for Alice}
\label{sec:efficient strategies}

We turn now to formalizing a strategy for Alice based only on a direct analysis of the current state of the game (without locating the state in the prefer-max sequence).  The idea is to analyze the states as numbers in base $k$. More precisely, we will consider the equivalence classes of states under cyclic rotation and rank them according to the highest number in base $k$ in a class. 
Note that this number can only increase when Bob plays. It decreases when Alice plays and kept constant when both players pass. We will show that Alice can play in a way such that Bob is forced to increase the value more than she decreases it, i.e., Alice can force the existence of a monotonically increasing subsequence of states. 
Therefore, the game will eventually reach the state $(k-1)^n$ (the state with the maximal value), from which Alice can play $n$ consecutive steps and reach her goal - the state $0^n$.

Towards this goal, we define the value of states as follows. The function $\val$ reads the state as a number in base $k$, and the function $\val^*$ assigns to each state the highest value in its equivalence class:
\begin{definition}\label{def:val_und_freund}
	For a state $s= \sigma_0\cdots\sigma_{n-1} \in [k]^n$ let $\val(s)=\Sigma_{i=0}^{n-1} \sigma_{n-1-i} \cdot k^{i}$ and 
	$\val^*(s)=\max\val(yx) : s = x y \}$.
\end{definition}

\begin{example}\label{example:val_y_amigos}
	    $\val(120)=0{\cdot} 1+2{\cdot} 3 +1{\cdot} 3^2 = 15$,         
	    $\val(201)=1{\cdot}1 +0 {\cdot} 3 +2 {\cdot} 3^2 =19$ and
        $\val(021)=1{\cdot} 1 +2{\cdot} 3 +0{\cdot} 3^2 =7$, 
        so $\val^*(120)=\max\{15,19,7\}=19$.
\end{example}

Definition~\ref{def:val_und_freund} is related to the known notion of \emph{Lyndon words}~\cite{lyndon1954burnside,berstel2007origins} - non-periodic words  that are lexicographically least among their rotations. Specifically, considering the function $\fneg\colon [k]^+ \to [k]^+$ defined by $\fneg(\sigma_0\cdots\sigma_{l})=(k-1-\sigma_0) \cdots (k-1-\sigma_{l})$, the relation is: the state  $\fneg(s)$  is a Lyndon word if and only if it is non-periodic and $\val^*(s)=\val(s)$.

Based on Definition~\ref{def:val_und_freund}, we propose the following strategy for Alice:
\begin{definition}\label{def:adagger}
Let $A^\dagger \colon [k]^n \to \{0,1\}$  be the strategy for Alice defined by
	$$
		A^\dagger (0^lw\sigma) =
		\begin{cases}
			1 & \text{if } \sigma \neq 0 \text{ and } \val^*(0^lw\sigma)=\val(w \sigma 0^l);\\
			0 & \text{otherwise}.
		\end{cases}
	$$
for any word $w$ that doesn't start with $0$.
\end{definition}

It is easy to see that $A^\dagger$ is a strategy for Alice, i.e., $A^\dagger(x0)=0$ for every $x \in [k]^{n-1}$.

\begin{example}\label{example:dagger}

	$A^\dagger (120) = A^\dagger (012) = 0$, 	$A^\dagger (201) = 1$.
	
	Note that in the first and the third cases the number of leading zeros is $l=0$, whereas in the second $l=0$.
\end{example}


We will make use of the following monotony property of $A^\dagger$:
\begin{proposition}
\label{prop:ADaggerMonotony}
For any $x \in [k]^{n-1}$ and $\sigma \in [k] \setminus \{0\}$, $A^\dagger(x\sigma) \geq A^\dagger(x\tau) $ for all $\tau > \sigma$.

\end{proposition}
\begin{proof}
Let $x = 0^lw$ where the first symbol in $w$ is not $0$. We need to show that if $\val(w\tau0^l)=\val^*(w\tau 0^l)$, then $\val(w \sigma 0^l)=\val^*(w \sigma 0^l)$. To this end, we will show that the value of any rotation of $x \sigma 0^l$ is not greater than $\val(x \sigma 0^l)$. The claim clearly holds for rotations that start with $0$, thus we consider a partition $x=y_1y_2$, and a rotation $y_2 \sigma 0^{l}y_1$.
By assumption, $\val(y_2\tau0^ly_1)\leq\val(y_1y_2\tau 0^l)$, i.e., $\Delta_\tau = \val(y_1y_2\tau 0^l) - \val(y_2\tau0^ly_1) \geq 0 $.
We turn to analyze  $\Delta_\sigma = \val(y_1y_2 \sigma 0^l) - \val(y_2 \sigma 0^l y_1)$, as required,
we get, by the definition of $\val$,  that $\Delta_\sigma=\Delta_\tau + (\tau-\sigma)(k^{l+|y_1|}  - k^{l}) > 0$  which means that $\val(y_2 \sigma 0^l y_1) <   \val(y_1y_2 \sigma 0^l)$ as required.
\end{proof}

A fact that will play a key role in Section~\ref{sec:efficientshiftrules} is that Definition~\ref{def:adagger} is related to the predicate $\head$  given in Definition 2 in~\cite{amram2017efficient}) as follows:

$$
		\head((k{-}1)^l w\sigma) =
		\begin{cases}
			true & \mbox{if  $\sigma \neq k-1$ and $w \sigma (k-1)^l$ and is} \\ 
                    & \mbox{lexicographically minimal among its rotations;}\\
			false & \text{otherwise}.
		\end{cases}
	$$
Specifically, the relation is:
\begin{proposition}
	\label{reinforcement}
	$\head(\fneg(s)) \Leftrightarrow A^\dagger(s)=1$.
\end{proposition}
\begin{proof}
$A^\dagger(s)=1\Leftrightarrow s=0^l w\sigma$ where $\sigma \neq 0$ and $w\sigma0^l$ is lexicographically largest among $s$'s rotations $\Leftrightarrow \fneg(s)=(k-1)^l \fneg(w)\fneg(\sigma)$ where  $\fneg(\sigma)\neq k-1$, and $\fneg(w)\fneg(\sigma)(k-1)^l$ is lexicographically smallest among $\fneg(s)$'s rotations $\Leftrightarrow \head(\fneg(s))$. 
\end{proof}


\section{$A^\dagger$ is a non-losing efficiently computable strategy for Alice}
\label{sec:alsiisas}
In this section we propose an algorithm for efficient computation of the $A^\dagger$ strategy for Alice, and show that it is a non-losing strategy. Using the uniqueness of  $A^*$, this will lead us to the conclusion that $A^\dagger=A^*$, therefore we can efficiently compute $A^*$. Moreover, by Proposition~\ref{proposition:pmaxeq},  the same applies to $B^*$.

\begin{proposition}\label{prop:O(n)_not_to_market}
	$A^\dagger(s)$ can be computed in $O(|s|)$ time and memory.
\end{proposition}
\begin{proof}
Let $s=\sigma w$ where $\sigma \in [k]$ and $w \in [k]^n$.
In order to compute $A^\dagger(s)$, we are to find $s^* \in \arg\max\{\val(s') \colon s' \text{ is a rotation of } s\}$.
We can compute the value of $\val(\sigma x)$ based on $\val(x \sigma )$ in $O(1)$ time and memory using the equation $\val(\sigma x)= (\val(x\sigma )-\sigma)/k +\sigma \cdot k^n$. Therefore, we can extract $s^*$ in $O(n)$ time and memory. Now, given $s^*$, we can compute $A^\dagger(s)$ in $O(n)$ time and space using:
$A^\dagger(s) =1 \iff s^* = 0^lw\sigma \text{ and } s=w\sigma0^l \text{ where } \sigma \neq 0$.
\end{proof}


The proof of the following proposition is a mathematical formulation of the intuition stated before the definition of $A^\dagger$. Specifically, it explains in details how Alice forces Bob to increase $\val^*$ more than she decreases it and how this drives the game, if Bob plays optimally, to the state $(k-1)^n$ from which Alice wins in $n$ steps. 

\begin{proposition}\label{proposition:lsiAwins}
$A^\dagger$ is a non-losing strategy for Alice.
\end{proposition}

\begin{proof}
We consider an infinite state progression $s_0,s_1,\dots$ when Alice plays according to $A^\dagger$, Bob plays a strategy $B$, and $s_0=0^{n}$. We show that Bob does not win, regardless of the chosen strategy $B$. Bob wins against $A^\dagger$ iff, excluding $s_0$, $0^n$ does not appear in the sequence. To show that $0^n=s_i$ for some $i>0$, we assume otherwise, and prove the following:
\begin{quote}
    There exists a subsequence of the state-progression, $s_{i_0},s_{i_1},\dots$ such that $\val^*(s_{i_0})<\val^*(s_{i_1})<\cdots$.
\end{quote}
Clearly, this leads to a contradiction as $\val^*$ cannot infinitely grow.

We start with a few technical claims.

\begin{claim}
\label{clm:ADaggerMonotony}
If $\val(x\tau0^l)=\val^*(x\tau 0^l)$, then $\val(x00^l)=\val^*(x0 0^l)$.
\end{claim}
\begin{proof}[proof of Claim~\ref{clm:ADaggerMonotony}]
Repeat the argument in the proof of Proposition~\ref{prop:ADaggerMonotony} for the case $\sigma=0$. \qedhere{Claim~\ref{clm:ADaggerMonotony}}
\end{proof}

\begin{claim}
\label{clm:lsi-add-zeros}
Let $t_1<t_2$ be such that $A^\dagger(s_{t_1})=1$, $t_2-t_1\leq n$, and $B(s_t)=0$  for every $t$ in range $t_1\leq t<t_2$. Write $s_{t_1}=xy$ where $|y|=t_2-t_1$. Then, $s_{t_2}=0^{t_2-t_1}x$.
\end{claim}
\begin{proof}[proof of Claim~\ref{clm:lsi-add-zeros}]
 Let $w(\tau+1)0^l$ be the rotation of $s_{t_1}$ that satisfies $\val^*(s_{t_1})=\val(w(\tau+1)0^l)$.
Therefore, as $A^\dagger(s_{t_1})=1$, $s_{t_1}=0^l w(\tau+1)$. Write $s_{t_1}=xy$ such that $|y|=t_2-t_1$. First, we prove only for the case $|y|\leq n-l$. Hence, we can write $s_{t_1}=xy = 0^lzy$. Recall that $\val^*(s_{t_1}=0^lzy)=\val(w(\tau+1)0^l=zy0^l)$ and hence, $\val^*(zy0^l)=\val(zy0^l)$. Write $y=\sigma_m\sigma_{m-1}\cdots\sigma_1$. We prove by induction that for $i\in\{0,1,\dots,m\}$, $s_{t_1+i}=0^{l+i}z\sigma_m\cdots\sigma_{i+1}$.

The induction hypothesis vacuously holds for $i=0$. For the induction step, assume that $s_{t_1+i}=0^{l+i}z\sigma_m\cdots\sigma_{i+1}$. If $\sigma_{i+1}=0$, the claim holds by the definition of a strategy for Alice. Otherwise,  as $\val(z\sigma_m\cdots\sigma_10^l)=\val^*(z\sigma_m\cdots\sigma_10^l)$, by Claim~\ref{clm:ADaggerMonotony}, also $\val(z\sigma_m\cdots\sigma_{i+1}0^{l+i})=\val^*(z\sigma_m\cdots\sigma_{i+1}0^{l+i})$. Consequently,  $A^\dagger(0^{l+i}z\sigma_m\cdots\sigma_{i+1})=1$ and thus $s_{t_1+i+1}=0^{l+i+1}z\sigma_m\cdots\sigma_{i+2}$. As a result, $s_{t_2}=s_{t_1+|y|}=0^{l+|y|}z=0^{|y|}x$, as required. 

Now, the case $|y|>l$ is argued as follows: As proved so far, $s_{t_1+(n-l)}=0^n$. Hence, as long Bob outputs $0$, the state remains $0^n$.\qedhere{Claim~\ref{clm:lsi-add-zeros}}
\end{proof}

\begin{claim}
\label{clm:2n-steps}
 Let $t_1$ be such that for every $t$ in range $t_1\leq t< t_1+2n$, $B(s_t)=0$. Then, for some $t$ in that range, $s_t=0^n$.
\end{claim}
\begin{proof}[proof of Claim~\ref{clm:2n-steps}]
If $s_{t_1}=0^n$ we are done, and otherwise, let $w(\sigma+1)0^l$ be the rotation of $s_{t_1}$ that satisfies $\val^*(s_{t_1})=\val(w(\sigma+1)0^l)$. Hence, $0^lw(\sigma+1)$ is the only rotation of $s_{t_1}$ on which $A^\dagger$ outputs $1$. As Bob keeps passing the turn to Alice, after $i$-steps for some $i<n$, the state progression reaches a state $s_{t_1+i}=0^lw(\sigma+1)$. Hence, by Claim~\ref{clm:lsi-add-zeros}, $s_{t_1+i+n}=0^n$.    
\qedhere{Claim~\ref{clm:2n-steps}}
\end{proof}

\begin{claim}
 \label{clm:Bob-plays-twice}
 If $t_1<t_2$ are such that $B(s_{t_1})=B(s_{t_2})=1$ and $s_t \neq 0^n$ for all $t_1\leq t\leq t_2$ then $\val^*(s_{t_1+1})<\val^*(s_{t_2+1})$.
\end{claim}
\begin{proof}[proof of Claim~\ref{clm:Bob-plays-twice}]
Note that it is sufficient to prove the claim for the case in which $B(s_t)=0$ for every $t$ in range $t_1<t<t_2$. 

First, assume that for every $t$ in range $t_1<t<t_2$, $A^\dagger(s_t)=0$. In this case, $s_{t_2}$ is a rotation of $s_{t_1+1}$. As Bob increases one of the symbols of $s_{t_2}$, the claim follows from the definition of $\val^*$,

Now, assume that for some intermediate state $s_t$ where $t_1<t<t_2$, $A^\dagger(s_t)=1$, and take such minimal $t$. Therefore, $s_t$ is a rotation of $s_{t_1+1}$ and thus
\begin{equation}
\label{eq:t-like-t1}
\val^*(s_{t_1+1})=\val^*(s_t).
\end{equation}

Let $l \in [n]$, $w \in [k]^{n-l-1}$ , and $\sigma \in [k] \setminus \{0\}$  be such that $w \sigma 0^l$ is the rotation of $s_{t_1+1}$ that satisfies $\val^*(s_{t_1 +1})=\val(w \sigma 0^l)$. Hence, as $A^\dagger(s_{t})=1$, according to the definition of $A^\dagger$, $s_t=0^l w \sigma$. By Claim~\ref{clm:lsi-add-zeros}, there are $x,\tau,y$ such that $w=x \tau y$  and $s_{t_2}=0^{l+1 + |y|} x \tau$. Therefore, $s_{t_2+1}=(\tau+1)0^{l+1+|y|} x$.

Now, $\val^*(s_t)=\val(w \sigma 0^l) = \val(x \tau y \sigma 0^l)$. As a result,
\begin{equation}
    \label{eq:increase-after-0s}
    \val^*(s_{t_2+1}) 
    \geq \val(x (\tau+1) 0^{l+1+|y|})
    >\val(x \tau y \sigma 0^l)
    =\val^*(s_t).
\end{equation}
The first inequality is by the definition of $\val^*$ and the fact that $x(\tau+1) 0^{l+1+|y|}$ is a rotation of $s_{t_2+1}$. 
The second inequality can be understood if one thinks of $\val$ as reading the word in base $[k]$ : we decreased a significant digit and increased less significant ones.  

Equations~\ref{eq:t-like-t1} and~\ref{eq:increase-after-0s} imply that $\val^*(s_{t_1+1})<\val^*(s_{t_2+1})$.\qedhere{Claim~\ref{clm:Bob-plays-twice}}
\end{proof}
We can finally prove the proposition. Assume towards contradiction that Bob wins with a strategy $B$. Hence, the obtained play $(s_0,\dots,s_m)$ does not include $0^n$, excluding $s_0$, and $s_m=s_r$ for some $0<r<m$. Consider the infinite state progression from $s_0=0^n$, $(s_i)_{i=0}^\infty$. As Bob wins, excluding $s_0$, this sequence does not include $0^n$. Hence, by Claim~\ref{clm:2n-steps}, there are infinitely many indices $i$  for which $B(s_i)=1$. Therefore, by Claim~\ref{clm:Bob-plays-twice}, there exists an infinite subsequence $(s_{i_j})_{j=1}^\infty$ such that $\val^*(s_{i_1})<\val^*(s_{i_2})<\cdots$, in contradiction to the fact that $\val^*(w)$ is bounded by $k^n$.
\end{proof}

\begin{proposition}\label{proposition:lsiA}
	$A^\dagger=A^*$.
\end{proposition}
\begin{proof}
	From Proposition~\ref{proposition:lsiAwins}, we have that $A^\dagger$ is a non-losing strategy, and from Proposition~\ref{proposition:AUnique} we know that $A^{*}$ is the only non-losing strategy for Alice.
\end{proof}


\section{Efficiently computable shift rules}
\label{sec:efficientshiftrules}
In this section we apply the efficient strategies developed above to propose an efficiently computable shift rules for both prefer min and prefer max De Bruijn sequences, in both directions (backwards and forwards).

We begin with an efficiently computable shift rule for the reverse of the prefer-max cycle:

\begin{theorem} \label{thm:main}
	The function
	$$ \shift(x \sigma)=
		\begin{cases}
			(\sigma+1) x & \text{if $\sigma < k-1$ and $A^\dagger(x(\sigma+1))=1$}; \\
			0x           & \text{else, if $A^\dagger(x\sigma)=1$};                  \\
			\sigma x     & \text{otherwise}.
		\end{cases}	$$
	maps each vertex on the prefer-max cycle to its predecessor and can be computed in $O(n)$ time and memory.
	
\end{theorem}

\begin{proof}
From  Proposition~\ref{proposition:lsiA}, we know that $A^\dagger=A^{*}$ and by Proposition~\ref{proposition:pmaxeq}, $A^\dagger(x (\sigma+1))=B^{*}(x \sigma)$.
By the definition of the game (Definition~\ref{def:shift-game}) we have that $s_{t+1}=\shift(s_t)$ where  $(s_t)_{t=0}^{k^n-1}$ is the play of the $(n,k)$-shift-game when Alice uses the $A^{*}$ strategy and Bob uses the $B^{*}$ strategy.
From Observation~\ref{observation:SisW} we get that 
	$(s_t)_{t=0}^{k^n-1} = (w_{k^n-1-t})_{t=0}^{k^n-1}$, where $(w_{t})_{t=0}^{k^n-1}$ is the prefer-max cycle given in Definition~\ref{def:pref-max}.
    \end{proof}

We next state an efficiently computable shift rule for the prefer-max cycle in the forward direction:

\begin{theorem} 
\label{theorem:shift-1}
The function
	$$  \shift^{-1}(\sigma x) = \begin{cases}
			x (\sigma-1) & \text{if $\sigma > 0$ and $A^\dagger(x\sigma)=1$};                                             \\
			x (\max S)   & \text{if $\sigma=0$ and $S=\{ \tau \neq 0 \colon A^\dagger(x\tau)=1 \} \neq \emptyset$}; \\
			x \sigma     & \text{otherwise}.
		\end{cases}
    $$
is the inverse of $\shift$. It maps each vertex on the prefer-max cycle to its successor. 
\end{theorem} 
\begin{proof}

Since $\shift$ is a bijection, it is sufficient to show that $\shift^{-1}(\shift(x\sigma)) = x\sigma$.  Let $\sigma' x = \shift(x \sigma)$.
    We split the proof into three cases, in correspondence with the three cases of $\shift$:
	\begin{enumerate}
		\item If $\sigma < k-1$ and $A^\dagger(x(\sigma+1))=1$:  Since $\shift(x\sigma)=(\sigma+1)x$ we can write $\sigma' = \sigma+1$.
		Thus $\sigma'>0$ and $A^\dagger(x\sigma')=1$, we fall in the first case when computing $\shift^{-1}(\sigma' x)$ , and get that  $\shift^{-1}(\sigma'x) = x(\sigma'-1)= x\sigma$.
		
		\item If 
		\begin{enumerate}
		    \item $\sigma = k-1$ or $A^\dagger(x(\sigma+1)) = 0$; and
		    \item $A^\dagger(x\sigma)=1$,
		    \end{enumerate}
		    by the definition of $\shift$, $\sigma'=0$.  Thus, the first case clause of  $\shift^{-1}(\sigma' x)$ is false. 
		We next show that we satisfy the conditions for the second case of $\shift^{-1}$, and that $\max S = \sigma$. By (b), we have that $\sigma \in S$, thus $S$ is not empty and we are in the second case of $\shift^{-1}$. If $\sigma = k-1$, it is the maximal value in $S$. Otherwise, by (a), 	$\sigma+1 \notin S$ thus, by Proposition~\ref{prop:ADaggerMonotony}, $\sigma$ is the maximum of $S$. Therefore, $\shift^{-1}(\sigma'x) = x\sigma$.

		\item If 
		\begin{enumerate}
		    \item $\sigma = k-1$ or $A^\dagger(x(\sigma+1)) = 0$; and
		    \item $A^\dagger(x\sigma) = 0$,
		    \end{enumerate}
		by the definition of $\shift$, $\sigma'=\sigma$.
		We claim that the first two cases of $\shift^{-1}(\sigma x)$ are both false. 
		First, by (b), the first case of $\shift^{-1}(\sigma x)$ is false.
		As for the second case, we split the proof to two:
		If $\sigma>0$, obviously the second case of $\shift^{-1}(\sigma x)$ is false. When $\sigma=0$, we are to show that $S=\emptyset$. By the definition of $A^\dagger$,  $0\notin S$.
		By (a), and as we assumed $\sigma=0$, we have that $A^\dagger(x(\sigma+1)) = 0$, thus $1 \notin S$ and by Proposition~\ref{prop:ADaggerMonotony}, $S = \emptyset$.
		Therefore, we are in the third case of $\shift^{-1}$, and thus $\shift^{-1}(\sigma x) = x\sigma$. \qedhere
		
	\end{enumerate}

\end{proof}

The above two definitions describe shift rules for the prefer-max sequence. For the prefer-min sequence, we can apply $\fneg$ on the vertices of the $(n,k)$-prefer-max cycle to get the $(n,k)$-prefer-min cycle (and vice verse). Using this, the next two corollaries gives an efficiently computable shift rule for the prefer-min cycle, both in the backward and forward directions:

\begin{corollary}
\label{cor:neg_shift_neg}
$\fneg(\shift(\fneg(s)))$ maps each vertex on the prefer-min cycle to its predecessor. It can be computed in $O(n)$ time and memory.
\end{corollary}

\begin{corollary}
\label{cor:neg_shift-1_neg}
$neg(\shift^{-1}(neg(s)))$ maps each vertex on the prefer-min cycle to its successor.
\end{corollary}

The complexity analysis of the forward direction depends on our ability to efficiently compute $\fneg(\shift^{-1}(\fneg(s)))$. As we show in the following proposition, this function was already analyzed by Amram et. al.~\cite{amram2017efficient} under the name $next$.

\begin{proposition}
    
 \label{thm:next}
The function  
$$next(\sigma x)=\begin{cases}
		x (\sigma+1) & \text{if $\sigma \neq k-1$ and $\head(x\sigma)$};                                            \\
		x (\min S)   & \text{else, if $\sigma=k-1$ and $S=\{ \tau \neq k-1 \colon \head(x\tau) \} \neq \emptyset$}; \\
		x \sigma     & \text{otherwise},
	\end{cases}$$ 
	defined by Amram et al.~\cite[Definition~3]{amram2017efficient}, satisfies $$next(\sigma x) =\fneg(\shift^{-1}(\fneg(\sigma x))).$$
\end{proposition} 
\begin{proof}

Let $\hat{\sigma}=\fneg(\sigma)$, $\hat{x}=\fneg(x)$ and $\hat{S}=\{ \hat{\tau} \neq k-1 \colon \head(\hat{x}\hat{\tau}) \}$. 
	Note that, by Proposition~\ref{reinforcement}, $\hat{S}=\{ \hat{\tau} : \tau \in S \}$.
	We need to show the following:

	\begin{itemize}
		\item $\sigma > 0 \wedge A^\dagger(x\sigma)=1$ if and only if $\hat{\sigma} \neq k{-}1 \wedge \head(\hat{x}\hat{\sigma})$.
		\item $\sigma=0$ and $S \neq \emptyset$ if and only if $\hat{\sigma}=k{-}1$ and $\hat{S}\neq \emptyset$.
		\item $x(\sigma-1) = \fneg(\hat{x}(\hat{\sigma}+1))$.
		\item $x (\max S) = \fneg(\hat{x} (\min \hat{S}))$. 
		\item $x\sigma = \fneg(\hat{x}\hat{\sigma})$.
	\end{itemize}
	All of these are true by Proposition~\ref{reinforcement} and by the definition of $\fneg$.
\end{proof}

Since Amram et al.~\cite[Theorem~12]{amram2017efficient} proved that the complexity of $next$ is $O(n)$ in both time and memory, we can conclude that:
\begin{corollary}
    The shift rules given in Theorem~\ref{theorem:shift-1} and Corollary~\ref{cor:neg_shift-1_neg} for the forward direction of both the prefer-max and the prefer-min can be computed in $O(n)$ time and memory.
\end{corollary}

\section{A proof of the theorem of Fredricksen, Kessler, and Maiorana}
Beyond providing efficiently computable rules, our results also cater for a very simple proof of a well known theorem, as follows.

Let $L_0,L_1,\dots, L_m$ be a lexicographic enumeration of all Lyndon words whose length divides $n$.
The main result of~\cite{FreMai78} (rephrased to simplify the presentation) is:

\begin{theorem}[FKM]
$L_0 L_1 \cdots L_{m}$ is the $(n,k)$-prefer-min sequence.
\label{thm:FM}
\end{theorem}


This theorem was first proposed by Fredricksen, Kessler, and Maiorana~\cite{FreMai78} with a partial proof (only that $L_0 L_1 \cdots L_{m}$ is a de-Bruijn sequence, not that it is the prefer-min sequence). After more than 25 years, Eduardo Moreno provided an alternative proof~\cite{moreno2004theorem} and then, after an additional ten years, provided more details to this proof with Dominique Perrin and added a proof of the other part of the statement (that this is the prefer-min sequence)~\cite{moreno2015corrigendum}.
Here we show another proof of this theorem using the combinatorial game studied in this paper and the following known result:
\begin{theorem}[Amram et al.~\cite{amram2017efficient}]
\label{thm:fromamram}

$next$ is a shift rule for $L_0 L_1 \cdots L_{m}$.

\end{theorem}

Theorem~\ref{thm:fromamram} is stated and proved by Amram et al.~\cite{amram2017efficient} inside the proof of Theorem~4.  Finally, the proof of the FKM theorem follows immediately:
\begin{proof}[Proof of Theorem~\ref{thm:FM}]
By Proposition~\ref{thm:next} and Theorem~\ref{thm:fromamram}, $next$ is a shift rule for both the prefer-min cycle and for the concatenation of the Lyndon words.
\end{proof}


\bibliographystyle{elsarticle-num}
\bibliography{bib}{}

\end{document}